\documentclass[12pt]{article}

\usepackage{epsf}
\usepackage{latexsym,euscript}
\usepackage[dvips]{graphicx}
\usepackage{amsmath}
\usepackage{amsfonts}
\usepackage{amssymb}

\begin{document}

\begin{center}
{\Large \textbf{Twist free energy and critical behavior of $3D$ $U(1)$ LGT
at finite temperature}}

\vspace*{0.6cm}
\textbf{O.~Borisenko\footnote{email: oleg@bitp.kiev.ua}, 
V.~Chelnokov\footnote{email: chelnokov@bitp.kiev.ua}}

\vspace*{0.3cm}
{\large \textit{N.N.Bogolyubov Institute for Theoretical
Physics, National Academy of Sciences of Ukraine, 03143 Kiev, Ukraine}}
\end{center}

\begin{abstract}
The twist free energy is computed in the Villain formulation of the $3D$ $U(1)$ lattice gauge 
theory at finite temperature. This enables us to obtain renormalization group equations describing 
a critical behavior of the model in the vicinity of the deconfinement phase transition. These equations 
are used to check the validity of the Svetitsky-Yaffe conjecture regarding the critical behavior of the lattice
$U(1)$ model. In particular, we calculate the two-point correlation function of the Polyakov loops
and determine some critical indices. 
\end{abstract}

\section{Introduction} 

The critical behavior of pure lattice gauge theories (LGTs) at finite temperatures is well 
understood for non-abelian $SU(N)$ theories in various dimensions. In particular, the phase 
structure of a finite-temperature three-dimensional ($3D$) pure $SU(N)$ LGT with the standard 
Wilson action is thoroughly investigated both for $N=2,3$ and for the 
large-$N$ limit (see, {\it e.g.},~\cite{3D_sun} and references therein). The
transition is second order for $N=2,3$ and first order for $N>4$. In the case
of the $SU(4)$ gauge group, most works agree that the transition is weakly 
first order. The deconfining transition in $SU(N=2,3)$ LGTs belongs to the 
universality class of $2D$ $Z(N=2,3)$ Potts models. All these phase transitions are characterized 
by the spontaneous symmetry breaking of a $Z(N)$ global symmetry of the lattice action in the 
high-temperature deconfining phase. 

Surprisingly, the situation is much less clear for the $3D$ $U(1)$ LGT. 
The present state of affairs can be briefly summarized as follows. $3D$ theory was 
studied by Parga using Lagrangian formulation of the theory \cite{parga}. 
At high temperatures the system becomes effectively two-dimensional, in particular
the monopoles of the original $U(1)$ gauge theory become vortices of the $2D$ 
system. The partition function turns out to coincide (in the leading order of the 
high-temperature expansion) with the $2D$ $XY$ model in the Villain representation. 
The $XY$ model is known to have the Berezinskii-Kosterlitz-Thouless (BKT) 
phase transition of the infinite order \cite{berezin,kosterlitz1}.  
According to the Svetitsky-Yaffe conjecture the finite-temperature phase transition 
in the $3D$ $U(1)$ LGT should belong to the universality class of 
the $2D$ XY model \cite{svetitsky}. This means, firstly that the global $U(1)$ 
symmetry cannot be broken spontaneously because of the Mermin-Wagner 
theorem \cite{mwtheorem} and, consequently the local order parameter does not exist for this type of 
the phase transition. Secondly, the correlation function of the Polyakov loops (which become spins of the $XY$
model) decreases with the power law at $\beta \geq \beta_c$ implying a logarithmic potential between heavy electrons
\begin{equation}
P(R) \ \asymp \ \frac{1}{R^{\eta (T)}} \ ,
\label{PLhight}
\end{equation}
where the $R\gg 1$ is the distance between test charges.
The critical index $\eta (T)$ is known from the renormalization-group analysis
of Ref.\cite{kosterlitz1} and equals $\eta (T_c) =1/4$ at the critical point of the BKT
transition. For $\beta < \beta_c$, $t=\beta_c/\beta -1$ one has
\begin{equation}
P(R) \ \asymp \ \exp \left [ -R/\xi (t)  \right ] \ ,
\label{PLlowt}
\end{equation}
where the correlation length $\xi \sim e^{bt^{-\nu}}$ and the critical index $\nu=1/2$. 
Therefore, the critical indices $\eta$ and $\nu$ should be the same in the finite-temperature
$U(1)$ model if the Svetitsky-Yaffe conjecture holds in this case.
The first numerical check of these predictions was performed on the lattices 
$L^2\times N_t$ with $L=16, 32$ and $N_t=4,6,8$ in \cite{mcfinitet}. 
Though authors of \cite{mcfinitet} confirm the expected BKT nature of the phase transition, 
the reported critical index is almost three times larger of that predicted for the $XY$ model, 
$\eta \approx 0.78$. More recent analytical and numerical studies of Ref.\cite{beta_szero} 
indicate that at least on the anisotropic lattice in the limit of vanishing spatial coupling $\beta_s$
(where space-like plaquettes are decoupled) the $3D$ $U(1)$ gauge model exhibits the critical behavior 
similar to the $XY$ spin model. However, numerical simulations of the isotropic model on the lattices 
up to $L=256$ and $N_t=8$ reveal that $\eta\approx 0.49$, {\it i.e.} still far from the $XY$ 
value \cite{u1_isotropic}.  Thus, so far there is no numerical indications that critical indices 
of $3D$ $U(1)$ LGT coincide with those of the $2D$ $XY$ model and the question of the universality 
remains open if $\beta_s$ is non-vanishing. 

On the analytical side one should mention a renormalization group (RG) study of 
Refs.\cite{svetitsky,borisenko}. In both cases a high-temperature and a dilute monopole gas 
approximations were used for the Villain formulation which helped to derive an effective 
sine-Gordon model. Resulting RG equations were shown to converge rapidly with iterations 
to RG equations of the $2D$ $XY$ model. It gives a strong indication that, indeed the nature 
of the phase transitions in both models is the same. Moreover, since the scaling 
of the lattice spacing coincides in both cases the critical index $\nu$ should also be the same 
(this however was not proven). Furthermore, neither critical points nor index $\eta$ has been 
determined in previous studies.  

In this work we re-examine the critical behavior of the Villain formulation of the $3D$ $U(1)$ LGT 
aiming to compute both critical indices $\nu$ and $\eta$ as well as to determine the location 
of the critical points. In order to achieve this goal we calculate the free energy of the model 
in the presence of a twist and express it like a function of a bare coupling, a monopole activity 
and adimensional ratio of the anisotropic couplings. Varying the lattice cut-off one then finds 
the RG equations in a standard manner. We analyze the equations thus obtained for different 
values of $N_t$. Also, we present results for the correlation function of the Polyakov loops 
which allow to extract the index $\eta$ at the critical point.

\section{Definition of the model and its dual}

We work on a periodic $3D$ lattice $\Lambda = L^2\times N_t$ with spatial extension $L$ 
and temporal extension $N_t$.  We introduce anisotropic dimensionless couplings as 
\begin{equation}
 \beta_t = \frac{1}{g^2a_t}  \ , \;\;\;\;\; \beta_s = \frac{\xi}{g^2a_s} \ = \ 
\beta_t \ \xi^2 \ , \;\;\;\;\; \xi = \frac{a_t}{a_s} \ ,
\label{ancoupl}
\end{equation}
where $a_t$ ($a_s$) is lattice spacing in the time (space) direction, 
$g^2$ is the continuum coupling constant with dimension $a^{-1}$. 
$\beta = a_tN_t$ is an inverse temperature. 

The compact $3D$ $U(1)$ LGT on the anisotropic lattice in the presence of the twist 
is defined through its partition function as 
\begin{equation}
Z(\beta_t,\beta_s) = \int_0^{2\pi}\prod_{x\in\Lambda}\: \prod_{n=1}^3
\frac{d\omega_n (x)}{2\pi} \ \exp{S[\omega + \theta]} \ ,
\end{equation}
where $S$ is the Wilson action 
\begin{eqnarray}
\label{wilsonaction}
 S[\omega] &=& \beta_s\sum_{p_s} \cos\omega (p_s) + 
\beta_t\sum_{p_t} \cos\omega (p_t)  \  , \\ 
\omega (p) \ &=& \ \omega_n(x)\ + \ \omega_m(x+e_n) \ - \ 
\omega_n(x+e_m) \ - \ \omega_m(x) \ 
\label{plaq_angle} 
\end{eqnarray}
and sums run over all space-like ($p_s$) and time-like ($p_t$) plaquettes.
We take a constant shift $\theta_n$ on a stack of plaquettes wrapping 
around the lattice in the spatial directions, {\it e.g.} the shift $\theta_1$ on the plaquettes with 
coordinates $p=(n_2,n_3;x_1,0,0)$ and the shift $\theta_2$ on the plaquettes with 
coordinates $p=(n_1,n_3;0,x_2,0)$ (for a detailed description of the twist in LGT we refer the reader to 
Ref.~\cite{twist_gauge} where also some properties of the twisted partition function are discussed). 

In order to calculate the free energy in the presence of the twist we make the following quite 
standard steps:
\begin{itemize} 
\item  Perform duality transformations with the twisted partition function;

\item Replace the dual Boltzmann weight with the Villain formulation 
and calculate an effective monopole theory;

\item Sum up over monopole configurations in the dilute gas approximation.   

\end{itemize}
All these steps are well known in the context of the $3D$ $U(1)$ LGT and can be easily generalized for 
the anisotropic lattice in the presence of the twist. For the duality transformations we need an
approach of Ref.~\cite{u1dual_pbc} which takes correctly into account the periodic boundary conditions 
on the abelian gauge fields. For the anisotropic theory with twist we find 
\begin{equation} 
Z(\theta_n) \ = \ \sum_{h_n = -\infty}^{\infty} e^{i \sum_{n=1}^{2} \ h_n  \theta_n} \ Z(h_n) \ ,
\label{dualPF_twist}
\end{equation}
where the global summation over $h_n$ enforces the global Bianchi constraint on the periodic system and 
$Z(h_n)$ is the dual partition function 
\begin{equation}
 Z(h_n) \  = \ \sum_{r(x)=-\infty}^{\infty} \  
\prod_x \ \prod_{n=1}^{3} \ I_{r(x)-r(x+e_n)+\eta_n(x)}(\beta_n) \ . 
\label{Zh}
\end{equation}
Here $I_r(x)$ is the modified Bessel function and we have introduced sources $\eta_n(x)=\eta (l)$ as 
\begin{equation}
 \eta (l) \ = \ 
\begin{cases}
 h_n, & l\in P_d \ ,  \\
0, & \text{otherwise} \ ,
\end{cases}
\label{source}
\end{equation}
where $P_d$ is a set of links dual to twisted plaquettes (this set forms a closed loop on the dual lattice), 
$\beta_n=\beta_s, n=3$ and $\beta_n=\beta_t, n=1,2$.
In the limit $\beta_s=0$ and in the absence of the twist the partition function (\ref{dualPF_twist}) 
reduces to ($x=(x_1,x_2)$ runs now over two-dimensional lattice $L^2$)
\begin{equation} 
Z(0) \ = \ \sum_{r(x)=-\infty}^{\infty} \  
\prod_x \ \prod_{n=1}^{2} \ I_{r(x)-r(x+e_n)}^{N_t}(\beta_n) \ .
\label{betas_0}
\end{equation} 
In this limit the model becomes a generalized version of the $XY$ model, and it was studied both analytically 
and by Monte-Carlo simulations in Ref.\cite{beta_szero}. The firm conclusion of Ref.\cite{beta_szero} was 
that the model (\ref{betas_0}) is in the same universality class as the $XY$ model. 
Here we are going to study an opposite limit, namely $\beta_t > \beta_s\gg 1$ which lies close to the continuum 
limit of the full $3D$ $U(1)$ model.  
When both couplings are large it is customary to use the Villain approximation, {\it i.e.}
\begin{equation}
I_r(x)/I_0(x) \ \approx \ \exp \left ( - \frac{1}{2x} r^2  \right )  \ .
\label{PTVillainDef}
\end{equation} 
This dual form of the twisted partition function, Eqs.~(\ref{dualPF_twist})-(\ref{PTVillainDef}), is a starting point 
of the analysis in the next Sections.

\section{Free energy of a twist}

Substituting (\ref{PTVillainDef}) into the partition function (\ref{Zh}) we use the Poisson summation 
formula to perform summation over $r(x)$ variables. The partition function is factorized in the product 
of the dual massless photon contribution and the contribution from the monopole configurations 
\begin{equation} 
Z(h_n) \ = \  Z_{ph} \ Z_m \ .
\label{PT_fact}
\end{equation} 
Taking into account the definition (\ref{source}) and performing summation over the lattice
we write these contributions in the presence of the twist as 
\begin{equation}
Z_{ph} \ = \ \exp \left[ -\frac{N_t}{2 \beta_t} \left( h_1^2 + h_2^2 \right) \right] \ ,
\label{Z_ph}
\end{equation}
\begin{equation} 
Z_m \ =  \ \sum_{\{m_x\}} \exp \left[ -\pi^2 \sum_{x,x'} m_x G_{x x'} m_{x'} 
- \frac{2 \pi i}{L}\sum_x m_x \left( h_1 x_1 + h_2 x_2 \right) \right] \;.
\label{twisted_pf}
\end{equation}
Here, $G_{x x'}$ is the three-dimensional Green function on anisotropic lattice. 
For our purposes it is convenient to present it in the form ($x_3=t$)
\begin{equation}
\label{GreenFunc}
G_{x,t;x',t'} \  =  \ \frac{\beta_t}{N_t} \left( G_{x,x'}^{2d} + 
\sum_{k=1}^{N_t-1} e^{\frac{2\pi i}{N_t} k (t-t')} G_{x,x'}^{2d}(M_k) \right) \ ,
\end{equation} 
where $G_{x}^{2d}$ is massless and $G_{x}^{2d}(M_k)$ massive $2D$ Green function with a mass 
\begin{equation}
M_k^2 \ = \ \beta_t/\beta_s (1 - \cos 2 \pi k /N_t)\;.
\label{M_k}
\end{equation}
Since massive Green functions are exponentially suppressed for $x \neq x'$ near the continuum limit like 
$\exp(-M_k R)$ we keep in the sum over temporal momenta $k$ in (\ref{GreenFunc}) only the terms with 
smallest $M_k$, corresponding to $k = 1, N_t -1$. Introducing notations $m_x = \sum_{t=0}^{N_t - 1} m_{x,t}$, 
$r_x^k = \sum_{t=0}^{N_t - 1} m_{x,t} \exp \frac{2 \pi i k t}{N_t}$ and keeping only leading contribution in the Taylor 
expansion of the terms with $x \neq x'$ we bring $Z_m$ to the following form 
\begin{eqnarray}
&& Z_m  =  \sum_{\{m_{x,t}\}} \exp -\frac{\pi^2 \beta_t}{N_t} \left( \sum_{x,x'} m_x G_{x x'}^{2d} m_{x'} 
+ \sum_{k=1}^{N_t - 1} \sum_{x} r_x^k G_0^{2d}(M_k) r_x^{-k} \right) \nonumber \\
&& \prod_{x \neq x'}  \left(1 - \frac{2 \pi^2 \beta_t}{N_t} r_x^1 G_{x x'}^{2d}(M_1) r_{x'}^{-1} \right) 
\exp \left[- \frac{2 \pi i}{L} \sum_{x} m_x \left( h_1 x_1 + h_2 x_2 \right)\right] \;. 
\label{zm2d}
\end{eqnarray}
Consider a set of $m_{x,t}$ variables at one spatial $x$ position.
Since all non-vanishing $r_x^k$ are suppressed by massive Green functions,
the dominant contribution arises from the following configurations: 1) $m_x = 0, r_x^k = 0$; 
2) $m_x = 0$, $r_x^k = \pm \left(1 - \exp \frac{2 \pi i k}{N_t} \right) \exp \frac{2 \pi i k \tau}{N_t}$; 
3) $m_x = \pm 1$, $r_x^k = \pm \exp \frac{2 \pi i k \tau}{N_t}$.
Since $G_{x}^{2d}$ diverges logarithmically in  the large-volume limit, only neutral configurations 
$\sum_x m_x=0$ contribute in this limit. If $\frac{\beta_t}{N_t}=T/g^2$ is large enough we can restrict 
ourselves only to leading contribution with $m_z = 1$, $m_{z^{\prime}} = -1$ and sum over $(z,z^{\prime})$. 
Summing up over all these configurations we finally obtain after a long algebra
\begin{equation}
Z_m  \ \approx \ \exp \left [ L^2 \ \sum_{z \neq 0} \exp \left[-\frac{2 \pi^2 \beta_t}{N_t} \ D(z)  
+ \frac{2 \pi i}{L} \left( h_1 z_1 +h_2 z_2 \right) \right] F(z) \right ] \ .
\label{zm_final}
\end{equation} 
The constant overall factor has been omitted. Here, $D(z)$ is the infrared-finite Green function 
whose asymptotics is given by $D(z) \ \asymp \ \frac{1}{\pi} \ \log (z_1^2+z_2^2)^{1/2} + \frac{1}{2}$.
If $F(z)=1$, the partition function (\ref{zm_final}) coincides with the vortex partition function 
of the $XY$ model in the presence of the twist. For the case of the finite-temperature $U(1)$ LGT
the function $F(z)$ reads 
\begin{equation}
F(z) \ = \  C_1+C_2 \left ( G_z^{2d}(M_1) \right)^2   \ .
\label{Fz}
\end{equation}
It incorporates two new contributions. The constant contribution 
\begin{equation}
C_1 \ = \ \left(\frac{N_t W_0}{1 + 2 N_t W_1}\right)^2 \ \left ( 1+  
 \frac{16 \pi^4 \beta_t^2}{N_t^2} U(1-U) \sum_{x \neq 0} \left ( G_x^{2d}(M_1) \right)^2 \right )      
\label{C1}
\end{equation}
renormalizes a monopole activity while the second one proportional to $C_2$
\begin{equation}
C_2 \ = \ \frac{8 \pi^4 \beta_t^2}{N_t^2} \ \left(\frac{N_t W_0}{1 + 2 N_t W_1}\right)^2 \ (1-U)^2 \ ,
\label{C2}
\end{equation}
gives an additional renormalization for the monopole-antimonopole logarithmic interaction at high temperatures. 
The constants introduced in the above equations are given by 
\begin{eqnarray*}
U &=& 2 (1 - \cos \frac{2 \pi}{N_t} ) 
\left( \frac{2 N_t W_1} {1 + 2 N_t W_1} \right)  \;, 
\label{sum_zm} \\
W_m &=& \exp \left [ -\frac{\pi^2 \beta_t}{N_t} \sum_{k=1}^{N_t - 1} 
\left(2-m-\cos \frac{2 \pi k}{N_t} m \right) G_0^{2d}(M_k) \right ] ,  m=0,1 \;. \nonumber   
\end{eqnarray*}
Noting that both $D(z)$ and $F(z)$ depend only on $r=(z_1^2+z_2^2)^{1/2}$ we can factorize the angular 
dependence of the twist. Integrating over the polar angle and replacing the summation over $r$ with 
integration near the continuum limit we find for the exponent of Eq.~(\ref{zm_final}) 
\begin{equation}
 L^2 \ \int_{1}^{+\infty} \exp \left[-\frac{2 \pi^2 \beta_t}{N_t} \ D(r) \right] F(r) \ 
J_0\left (\frac{2 \pi i}{L}r \right ) d r\ ,
\label{zm_exp}
\end{equation}   
where $J_0(x)$ is the Bessel function. Combining this result with Eq.~(\ref{Z_ph}) and 
summing up over $h_n$ in Eq.~(\ref{dualPF_twist}) gives us the following expression 
for the twisted partition function in the thermodynamic limit
\begin{equation}
Z(\theta) \  = \ \sum_{n_i=-\infty}^{+\infty} 
\exp {\left( -\frac{\beta_{eff}}{2} \sum_{i = 1,2}(\theta_i - 2 \pi n_i )^2 \right) } \ .
\label{Ztw_final}
\end{equation}
We have introduced here the renormalized coupling constant $\beta_{eff}$
\begin{equation}
\frac{1}{\beta_{eff}} \ = \ \frac{N_t}{\beta_t} + 
2 \pi^3 y^2 \int_{1}^{+\infty} r^{3- 2\pi \frac{\beta_t}{N_t}} \left( 1 + \frac{C_2}{C_1} \ \left( G_r^{2d}(M_1) \right)^2 \right)  d r \ . 
\label{tpart}
\end{equation} 
The first term corresponds to the massless photon contribution while the second one arises due to 
monopole-antimonopole interaction. The monopole activity $y$ is given by 
\begin{equation}
y \ = \ 2 \ C_1^{1/2} \exp \left({-\frac{1}{2} \pi^2 \frac{\beta_t}{N_t}}\right) \ .
\label{mon_y}
\end{equation} 
Following the same strategy one can compute the two-point correlation function of the Polyakov loops 
in the representation $j$ which appears to have a power-like decay of the form
\begin{equation}
 P_j(R) \ \approx \  \exp \left[ - \frac{j^2}{2 \pi \beta_{eff}} \ \ln R \right] \ .
\label{PL}
\end{equation}

\section{The renormalization group equations}

The RG equations can be derived from the expression for $\beta_{eff}$ by integrating 
in Eq.~(\ref{tpart}) between length scales $a$ and $a+\delta a$, see {\it e.g} \cite{nelson}. 
Renormalizing masses $M_k$ in such a way to preserve $G_r^{2d}(M_k)$ we obtain RG equations in 
a differential form as ($t=\ln a$)
\begin{eqnarray}
\frac{d \beta_t}{d t} &=& - 2 \pi^3 y^2 \frac{\beta_t^2}{N_t} \left( 1 + 
\frac{C_2}{C_1} \ \left( G_1^{2d}(M_1) \right)^2 \right)\ , \nonumber \\
\frac{d y}{d t} &=& y \; \left(2 - \pi \frac{\beta_t}{N_t}\right) \ , \ 
\frac{d M_k}{d t} = M_k \ .
\label{rg2}
\end{eqnarray}
When $N_t=1$ these equations are reduced to the equations of the $2D$ $XY$ model.
The equations for $M_k$ can be solved explicitly $M_k(t) = M_k(0) e^{t}$. Thus, 
$M_k$ grows exponentially with $t$ and in the limit $M_k \to \infty$ we again obtain 
RG equations of the $2D$ $XY$ model. 
Hence, we can expect that the critical indices of the model that describe the solution around a fixed point 
coincide with those of the $2D$ $XY$ model. 
To check that this is the case we solve the equations (\ref{rg2}) numerically in the vicinity 
of the fixed point $\beta_t = 2 N_t / \pi, \ y = 0$. 

\begin{figure}
\centering
\includegraphics[width=0.49\textwidth]{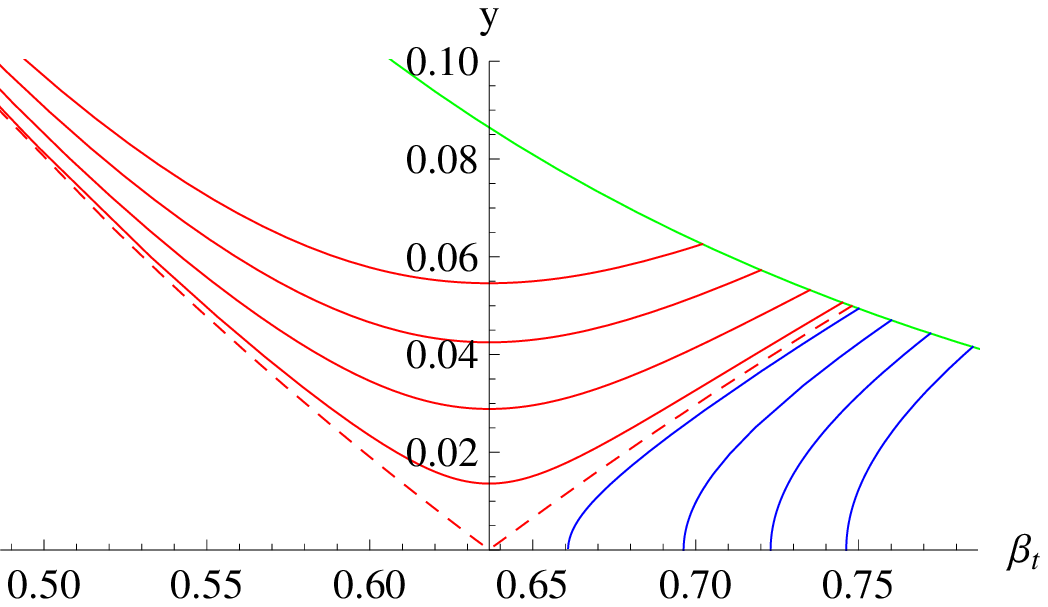}
\includegraphics[width=0.49\textwidth]{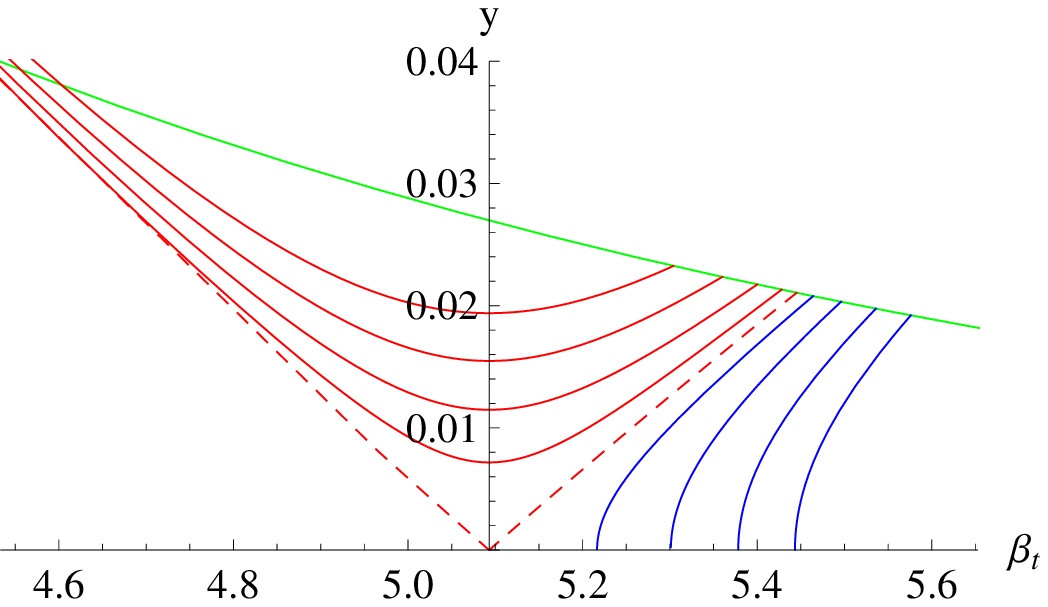}
\caption{Renormalization flow for $N_t = 1$ (left) and $N_t = 8$ (right) obtained from numerical solution 
of RG equations. Green line defines the initial points, dashed red line is the critical line, 
blue lines show RG flow in the massless phase ($\beta > \beta_{t,\rm crit}$),
red lines show RG flow in the massive phase ($\beta < \beta_{t,\rm crit}$).}
\label{rgflow}
\end{figure}

Fixing $\beta_s/\beta_t$ gives us an initial value for $M_k$. It should be sufficiently large to ensure 
the fast convergence of the Taylor expansion in Eq.~(\ref{zm2d}). We have studied several initial values 
and have found no difference in the final result. As an example, in the Fig.~\ref{rgflow} we compare the renormalization 
flow for $N_t = 1$ ($2D$ $XY$ model) with that of $N_t = 8$ taken $M_1 = 4$ as the initial value.  
The critical index $\nu$ can be obtained from fitting the values of the cut-off $a$ at which $\beta_t(a)$ flows 
to the fixed point $2 N_t / \pi$ from above (massive phase). As a fitting function we use  
$a \sim \exp A(\beta_t - \beta_{t,\rm crit})^{-\nu}$. 
Our results for the critical points and $\nu$ values are summarized in the Table~\ref{tb1_crit_betas}. 
We observe that for all $N_t$ the value of $\nu$ is compatible with the $XY$ value $\nu=1/2$. 
The critical index $\eta$ can also be determined at the fixed point. Since 
$\beta_{eff}(\beta_{t,\rm crit}) = 2/\pi$ we find from Eq.~(\ref{PL}) $\eta=1/4$ for $j=1$.

\begin{figure}
\centering
\includegraphics[width=0.49\textwidth]{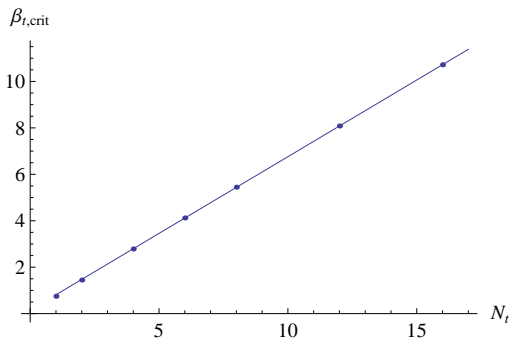}
\caption{Critical points for $N_t = 1,2,4,6,8,12,16$ fitted with the line 
$\beta_{t,\rm crit} = 0.139 + 0.661 N_t$.}
\label{betac_fit}
\end{figure}

To construct the continuum limit we fitted the critical couplings $\beta_{t,\rm crit}$ 
using several dependence on $N_t$. The best result is obtained with the fitting function  
$\beta_{t,\rm crit} = 0.139 + 0.661 N_t$. Thus, in the continuum limit the critical point 
is defined by $T_c \approx 0.661 g^2$. The Fig.~\ref{betac_fit} shows the fitting function 
together with values of $\beta_{t,\rm crit}$ from the Table~\ref{tb1_crit_betas}.

\begin{table}[ht]
\begin{center}
\begin{tabular}{|c|c|c|}
\hline
 $N_t$ & $\beta_{t,\rm crit}$ & $\nu$ \\
\hline
 1 & 0.748 & 0.498 \\   
 2 & 1.447 & 0.499 \\
 4 & 2.785 & 0.506 \\ 
 6 & 4.122 & 0.503 \\
 8 & 5.445 & 0.503 \\ 
12 & 8.082 & 0.504 \\
16 & 10.718 & 0.504 \\
\hline
\end{tabular}
\caption{Values of $\beta_{t,\rm crit}$ and $\nu$ obtained 
for various $N_t$.}
\label{tb1_crit_betas}
\end{center}
\end{table}

\section{Summary}

In this paper we have computed the twist free energy of the finite-temperature $3D$ $U(1)$ LGT 
in the Villain formulation. This enabled us to obtain and analyze the RG equations which 
describe the critical behavior of the model across the deconfinement phase transition.  
Our main findings can be shortly summarized as follows.
\begin{itemize}
 \item  We have computed the critical points for various temporal extension $N_t$. 
In the continuum limit we find  $T_c \approx 0.661 g^2$.

\item The scaling of the correlation length $\xi\sim a$ is compatible with a phase 
transition of the infinite order. Moreover, the critical index $\nu\approx 1/2$.

\item We have also derived the leading asymptotic behavior of the Polyakov loop 
correlation function. This allowed us to determine the critical index $\eta$ at 
the critical point $\eta (\beta_{t,\rm crit})=1/4$.
\end{itemize}
This supports the Svetitsky-Yaffe conjecture that the deconfinement phase transition 
in the finite-temperature $3D$ $U(1)$ LGT belongs to the universality class of the $2D$ $XY$ model, 
at least in the region of bare coupling constants where our approximations hold, {\it i.e.} 
for $\beta_t/\beta_s > 1$.  For isotropic lattices, used in \cite{u1_isotropic}, 
the initial value becomes $\beta_t/\beta_s = 1$. In this case one should take into account 
higher order terms of the Taylor expansion in the calculation of $Z_m$ which is hard to accomplish 
analytically. Still, we feel that the universality can be demonstrated also in this case 
by performing large-scale Monte-Carlo simulations of the isotropic model. 
Such computations are now in progress.

\end{document}